\documentclass[%
 reprint,
superscriptaddress,
 amsmath,amssymb,
 aps, prc
]{revtex4-2}
\usepackage{graphicx}
\usepackage{dcolumn}
\usepackage{bm}

\begin{document}

\title{Pairing correlations, orientations and quantum fluctuations in one- and two-nucleon transfer reactions at sub-barrier energies}

\author{D. D. Zhang}
\affiliation{Institute of Theoretical Physics, Chinese Academy of Sciences, Beijing 100190, China}

\author{B. Li}
\affiliation{State Key Laboratory of Nuclear Physics and Technology, School of Physics, Peking University, Beijing 100871, China}

\author{D. Vretenar}
\email{vretenar@phy.hr}
\affiliation{Physics Department, Faculty of Science, University of Zagreb, 10000 Zagreb, Croatia}
\affiliation{State Key Laboratory of Nuclear Physics and Technology, School of Physics, Peking University, Beijing 100871, China}

\author{T. Nik\v{s}i\'{c}}
\affiliation{Physics Department, Faculty of Science, University of Zagreb, 10000 Zagreb, Croatia}
\affiliation{State Key Laboratory of Nuclear Physics and Technology, School of Physics, Peking University, Beijing 100871, China}

\author{P. W. Zhao}
\email{pwzhao@pku.edu.cn}
\affiliation{State Key Laboratory of Nuclear Physics and Technology, School of Physics, Peking University, Beijing 100871, China}

\author{J. Meng}
\email{mengj@pku.edu.cn}
\affiliation{State Key Laboratory of Nuclear Physics and Technology, School of Physics, Peking University, Beijing 100871, China}

\date{\today}

\begin{abstract}
This work investigates one- and two-neutron transfer in the $^{96}\text{Zr} + {}^{40}\text{Ca}$ reaction at sub-barrier energies using a microscopic framework based on time-dependent covariant density functional theory (TD-CDFT). Pairing correlations are incorporated via the time-dependent BCS approximation, which is shown to significantly enhance pair transfer, as evidenced by an increased two-neutron transfer probability. The oblate deformation of $^{96}$Zr causes the transfer probabilities to vary by orders of magnitude with orientation; a direct comparison with experiment is enabled by averaging results over thirteen systematically chosen orientations. While the orientation-averaged one-neutron transfer probabilities agree well with data, the two-neutron channel is suppressed below the Coulomb barrier. This suppression is attributed to missing quantum fluctuations in the semiclassical TD-CDFT approach. To test this, we employ the generalized time-dependent generator coordinate method (TDGCM), which confirms that quantum fluctuations are essential for an accurate description of sub-barrier two-neutron transfer dynamics.
\end{abstract}

\maketitle

\section{Introduction}\label{sec1}

Historically, few-nucleon transfer reactions near the Coulomb barrier have been instrumental in advancing our understanding of nuclear structure and reaction dynamics. Processes involving the transfer of one or two nucleons are especially sensitive probes for nucleon-nucleon correlations, such as those arising from pairing.

Heavy-ion experiments -- including studies of $^{96}$Zr+$^{40}$Ca~\cite{Corradi2011PRC}, $^{116}$Sn+$^{60}$Ni~\cite{Montanari2014PRL, Corradi2022PLB}, $^{206}$Pb+$^{118}$Sn~\cite{Szilner2024PRL}, and $^{92}$Mo+$^{54}$Fe~\cite{Mijatovic2026PLB} -- have been conducted to elucidate the interplay between single-nucleon and pair-transfer mechanisms. A consistent result from these studies is that the probability for two-nucleon transfer is significantly enhanced over predictions from models of uncorrelated, sequential transfer. Specifically, the observed two-nucleon probability exceeds the square of its single-neutron counterpart, highlighting the role of strong correlations.

The interpretation of these results has spurred the development of diverse theoretical models, ranging from semiclassical time-dependent perturbation theory \cite{Oertzen2001RPP, Vitturi2012PTPS, Broglia1973PLB, Maglione1985PLB}, and the second-order distorted-wave Born approximation \cite{Potel2013RPP, Potel2011PRL, Potel2013PRC, Bayman1982PRC}, semiclassical \cite{Corradi2009JPG} to fully quantum-mechanical coupled-channels approaches \cite{Esbensen1989NPA, Esbensen1998PRC, Scamps2015PRC}.
Time-dependent density functional theory (TDDFT) provides a unified treatment of structure and dynamics, yet it exhibits a systematic flaw: it overestimates one-neutron transfer while underestimating the two-neutron process \cite{Scamps2015EPJ}. This unresolved discrepancy indicates an important topic for future research \cite{Abdurrahman2025}.

This study presents a comprehensive, microscopic investigation of one- and two-nucleon transfer, using the $^{96}$Zr+$^{40}$Ca system as a benchmark. Our analysis is structured in a multi-step approach to progressively incorporate critical physical effects. First, we employ time-dependent covariant density functional theory (TD-CDFT) -- a method well-established for modeling diverse nuclear dynamics like alpha scattering~\cite{Ren2020PLB}, fusion \cite{Ren2020PRC}, fission \cite{Ren2022PRC, Ren2022PRL, Li2023PRC, Li2024PRC}, quasifission~\cite{Zhang2024PRC-1}, and multinucleon transfer \cite{Zhang2024PRC, Li2024PRC, Zhang2025PLB} -- to analyze the primary reaction channels. To address the key role of nucleon correlations, we then augment this framework with dynamical pairing correlations via the time-dependent Bardeen-Cooper-Schrieffer (BCS) approximation~\cite{EbataPRC2010, ScampsPRC2013, Ren2022PRC, Ren2022PRL}. 
Furthermore, we account for the significant influence of nuclear deformation by sampling multiple orientations of the deformed $^{96}$Zr, a crucial factor in near-barrier reactions. While TD-CDFT captures the most probable pathways, a complete description of sub-barrier transfer requires quantum fluctuations. Therefore, we finally employ the generalized time-dependent generator coordinate method (TDGCM) \cite{Li2023PRC-1, Li2024FOP} to incorporate these effects, providing a more robust and complete theoretical description of the transfer process.

This paper is structured as follows. Section~\ref{sec2} outlines the theoretical framework, including the TD-CDFT formalism, the time-dependent BCS approximation, the generalized TDGCM, and the double-projection technique for particle number. Section~\ref{sec3} details the numerical methods for the static and dynamical calculations. The results are presented and discussed in Section~\ref{sec4}, with a focus on the roles of dynamical pairing, nuclear orientation, and quantum fluctuations. Finally, Section~\ref{sec5}  summarizes the principal conclusions.

\section{Theoretical Framework}\label{sec2}
\subsection{Time-dependent covariant density functional theory}
Nucleon transfer reactions are modeled using TD-CDFT with pairing correlations, treated dynamically in the time-dependent  BCS approximation. The wave function of the system takes the general form of a quasiparticle vacuum~\cite{EbataPRC2010, ScampsPRC2013, Ren2022PRC, Ren2022PRL},
\begin{equation}\label{eq: quasiparticle vacuum}
	|\Phi(\bm{r},t)\rangle=\prod_{k>0}[\mu_k(t)+\nu_k(t)c_k^{\dagger}(t)c_{\bar{k}}^{\dagger}(t)]|0\rangle,
\end{equation}
where $c_k^{\dagger}(t)$ and $c_{\bar{k}}^{\dagger}(t)$ denote the creation operators for the single-particle state $\psi_k(t)$ and the time-reversed state $\psi_{\bar{k}}(t)$, respectively. $\mu_k(t)$ and $\nu_k(t)$ are the transformation coefficients between the canonical and quasiparticle states.

The time evolution of the single-particle state $\psi_k(\bm{r},t)$ in coordinate space is governed by the Dirac equation~\cite{Ren2022PRC, Ren2022PRL},
\begin{equation}\label{eq: Dirac equation}
	i\hbar\frac{\partial}{\partial t}\psi_k(\bm{r},t)=[\hat{h}(\bm{r},t)-\varepsilon_k(t)]\psi_k(\bm{r},t).
\end{equation}
Here, $\hat{h}(\bm{r},t)$ is the single-particle Hamiltonian, and $\varepsilon_k(t)=\langle\psi_k(\bm{r},t)|\hat{h}(\bm{r},t)|\psi_k(\bm{r},t)\rangle$  denotes the single-particle energy. The point-coupling relativistic density functional PC-PK1~\cite{ZhaoPRC2010} is used in the following calculation, and the single-particle Hamiltonian $\hat{h}(\bm{r},t)$ can be written in the form
\begin{equation}
	\hat{h}(\bm{r},t)=\bm{\alpha}\cdot(\hat{\bm{p}}-\bm{V})+V^0+\beta(m+S),
\end{equation}
where
\begin{subequations}
	\begin{eqnarray}
		S(\bm{r},t)&=&\alpha_S\rho_S+\beta_S\rho_S^2+\gamma_S\rho_S^3+\delta_S\Delta \rho_S,\\
		V^{\mu}(\bm{r},t)&=&\alpha_Vj^{\mu}+\gamma_V(j^{\mu}j_{\mu})j^{\mu}+\delta_V\Delta j^{\mu}+\tau_3\alpha_{TV}j_{TV}^{\mu}\nonumber\\
		&+&\tau_3\delta_{TV}\Delta j_{TV}^{\mu}+e\frac{1-\tau_3}{2}A^{\mu}.
	\end{eqnarray}
\end{subequations}
$\bm{\alpha}, \beta$ are the Dirac matrices, $m$ is the mass of nucleon, and $\alpha_S, \alpha_V, \alpha_{TV}, \beta_S, \gamma_S, \gamma_V, \delta_S, \delta_V, \delta_{TV}$ are the coupling parameters. We refer the reader to Refs.~\cite{ZhaoPRC2010, Ren2020PRC} for further details. The scalar $S$ and vector $V^{\mu}$ potentials are determined by the time-dependent densities and currents, defined as follows:
\begin{subequations}
	\begin{eqnarray}
		\rho_S(\bm{r},t)&=&\sum_k n_k\bar{\psi}_k\psi_k,\\
		j^{\mu}(\bm{r},t)&=&\sum_k n_k\bar{\psi}_k\gamma^{\mu}\psi_k,\\
		j_{TV}^{\mu}(\bm{r},t)&=&\sum_k n_k\bar{\psi}_k\gamma^{\mu}\tau_3\psi_k,
	\end{eqnarray}
\end{subequations}
where $n_k(t)=\nu_k^2$ denotes the occupation probability of $k$-th single-particle state.

The time evolution of $n_k(t)=\nu_k^2(t)$, and the anomalous density $\kappa_k(t)=\mu_k^*(t)\nu_k(t)$, are governed by~\cite{Ren2022PRC, Ren2022PRL}
\begin{subequations}\label{eq7}
	\begin{eqnarray}
		i\frac{d}{dt}n_k(t)&=&\kappa_k(t)\Delta_k^*(t)-\kappa_k^*(t)\Delta_k(t),\\
		i\frac{d}{dt}\kappa_k(t)&=&[\varepsilon_k(t)+\varepsilon_{\bar{k}}(t)]\kappa_k(t)+\Delta_k(t)[2n_k(t)-1],
	\end{eqnarray}
\end{subequations}
respectively. The pairing gap $\Delta_k(t)$ is evaluated from 
\begin{equation}
	\Delta_k(t)=\left[G\sum_{k'>0}f(\varepsilon_{k'})\kappa_{k'}\right]f(\varepsilon_k),
\end{equation}
where $G$ is the strength parameter of the monopole pairing force, and $f(\varepsilon_k)$ is the cutoff function for the pairing window~\cite{ScampsPRC2013}. 

Although the time-dependent BCS approximation violates the one-body continuity equation, its effect is found to be moderate in the present analysis. For comparison, we also employ the frozen occupation approximation (FOA). In the FOA, pairing correlations are included in the ground state via the BCS method, but the occupation probabilities $n_k(t)$ and the anomalous density $\kappa_k(t)$ are fixed at their initial values throughout the time evolution.

\subsection{Generalized time-dependent GCM}

The TDGCM correlated nuclear wave function with discretized generator coordinates reads ~\cite{Regnier2019PRC, Li2023PRC-1, Marevicc2023PRC, Li2024FOP, Li2005PRC}
\begin{equation}
    |\Psi(t)\rangle=\sum_{\bm q} f_{\bm q}(t) |\Phi_{\bm q}(t)\rangle,
    \label{Eq_collec_wfs}
\end{equation}
where the vector ${\bm q}$ denotes {\em generator coordinates} that parametrize collective degrees of freedom.
This wave function is a linear superposition of, generally non-orthogonal, many-body {\em generator states} $|\Phi_{\bm q}(t)\rangle$ that have the form of Eq.~(\ref{eq: quasiparticle vacuum}), and their evolution in time is modeled by TD-CDFT. The equation of motion for the {\em weight functions} $f_{\bm q}(t)$ is obtained from the time-dependent variational principle~\cite{Regnier2019PRC},
\begin{equation}
    \sum_{\bm q} i\hbar \mathcal{N}_{\bm{q'q}}(t)\partial_t f_{\bm q}(t)=\sum_q \mathcal{H}_{\bm {q'q}}(t)f_{\bm q}(t)-\sum_q \mathcal{H}_{\bm {q'q}}^{MF}(t) f_{\bm q}(t)
    \label{TD-HW-f}
\end{equation}
where the time-dependent kernels
\begin{subequations}\label{eq: kernels}
 \begin{align}
&\mathcal{N}_{\bm{q'q}}(t)=\langle\Phi_{\bm {q'}}(t)|\Phi_{\bm q}(t)\rangle,\label{Eq_N}\\
&\mathcal{H}_{\bm{q'q}}(t)=\langle\Phi_{\bm {q'}}(t)|\hat{H}|\Phi_{\bm q}(t)\rangle,\label{Eq_H}\\
&\mathcal{H}^{MF}_{\bm{q'q}}(t)=\langle\Phi_{\bm {q'}}(t)|i\hbar\partial_t|\Phi_{\bm q}(t)\rangle,
\label{Eq_H_mf}
 \end{align}
\end{subequations}
include the overlap, the Hamiltonian, and the time derivative of the generator states, respectively.
 The corresponding collective wave function $g_q(t)$ is defined by the transformation~\cite{Reinhard1987RPP} 
  \begin{equation}
        g_{\bm{q'}}(t) = \sum_{\bm{q}} N^{1/2}_{\bm{q'}\bm{q}}(t) f_{\bm{q}}(t).
 \end{equation}
The expressions for the overlap kernel $\mathcal{N}_{\bm{q'q}}(t)$, the energy kernel $\mathcal{H}_{\bm{q'q}}(t)$, and the mean-field kernel $\mathcal{H}^{MF}_{\bm{q'q}}(t)$ in Eq.~(\ref{eq: kernels}) can be found in the Supplemental Material of Ref.~\cite{Li2005PRC}.

\subsection{Particle number projection: a double-projection technique} 

In TD-CDFT simulations with the time-dependent BCS approximation, the total particle number of the system is conserved on average. As nucleons transfer between the projectile and target, the collision produces two distinct fragments: a projectile-like fragment (PLF) and a target-like fragment (TLF). Once these fragments are fully separated, their individual wave functions are not eigenstates of the particle number operator. To determine the probabilities for one- and two-nucleon transfer channels, the double projection technique is therefore required~\cite{ScampsPRC2013}.

The coordinate space is divided into the region $V$, which contains the fragment we are interested in, and the complementary region $\bar{V}$. The probability that the fragment is composed of $m$ particles (protons or neutrons) can be calculated from
\begin{equation}\label{eq8}
	P_{m} = \frac{\langle\Psi|\hat{P}_m^V\hat{P}_M|\Psi\rangle}{\langle\Psi|\hat{P}_M|\Psi\rangle},
\end{equation}
with $M$ denoting the total number of particles (protons or neutrons) in the entire system. $\hat{P}_M$ and $\hat{P}_m^V$ are the particle number projection operators in the full space ($V+\bar{V}$) and subspace ($V$), respectively
\begin{subequations}\label{eq: pnp_operator}
	\begin{equation}
		\hat{P}_M=\frac{1}{2\pi}\int_0^{2\pi}d\theta e^{i\theta(\hat{N}-M)},
	\end{equation}
    \begin{equation}
    	\hat{P}_m^V=\frac{1}{2\pi}\int_0^{2\pi}d\theta' e^{i\theta'(\hat{N}_V-m)}.
    \end{equation}
\end{subequations}
$\hat{N}$ and $\hat{N}_V$ are the corresponding particle number operators in $V+\bar{V}$ and $V$, respectively
\begin{subequations}
\begin{equation}
	\hat{N}=\sum_{k}a_k^{\dagger}a_k,
\end{equation}
\begin{equation}
	\hat{N}_V=\sum_{k}\Theta_V(r_k)a_k^{\dagger}a_k,
\end{equation}
\end{subequations}
where the Heaviside function divides the coordinate space: 
\begin{equation}
\Theta_V(r_k)=\begin{cases}
		1,\quad r_k\in V,\\
		0,\quad r_k\notin V.
	\end{cases}
\end{equation}
We use the Pfaffian method~\cite{Regnier2019PRC} to calculate the overlap between the quasiparticle state and its gauge-angle-rotated state, both in the numerator and denominator in Eq. (\ref{eq8}). The specific implementation based on TD-CDFT used in the present work is described in the Supplemental Material of Ref.~\cite{Li2005PRC}.

Since the experimental one- and two-neutron transfer probabilities are determined by measuring the production probabilities of $^{41}{\rm Ca}$ and $^{42}{\rm Ca}$, respectively, the corresponding $x$-neutron transfer probability is calculated as
\begin{equation}\label{eq: Pxn}
	P_{xn} = P_{N_0+x}^{(n)}\, P_{Z_0}^{(p)},
\end{equation}
where $N_0$ and $Z_0$ are the neutron and proton numbers of the projectile $^{40}{\rm Ca}$, respectively, and $x$ is the number of transferred neutrons.

\section{Numerical Details}\label{sec3}

This study investigates one- and two-neutron transfer probabilities for the $^{96}{\rm Zr}+{}^{40}{\rm Ca}$ reaction using TD-CDFT and TDGCM. The theoretical results are compared with experimental data from Ref.~\cite{Corradi2011PRC}, which detected Ca-like recoils at a center-of-mass scattering angle of $\theta_{\text{c.m.}}=140^{\circ}$ using the PRISMA spectrometer.

The calculations are performed for a series of center-of-mass energies ($E_{\text{c.m.}}=76,~80,~84,~88,~92,~95$, and 96 MeV). Based on the classical relation between $E_{\text{c.m.}}$, $\theta_{\text{c.m.}}$, and the impact parameter $b$,
\begin{equation}
b=\frac{Z_{P}Z_{T}e^2}{2E_{\text{c.m.}}}\sqrt{\frac{1}{\sin(\theta_{c.m.}/2)^2}-1},
\end{equation}
these energies correspond to impact parameters of $b=2.76,~2.62,~2.50,~2.38,~2.28,~2.21$, and 2.18 fm, respectively.
Therefore, the calculated transfer probabilities correspond to peripheral collisions with $b\approx 2$-$3$ fm, consistent with the experimental conditions. In contrast, earlier TDDFT studies were typically restricted to central collisions ($b=0$ fm)~\cite{Scamps2015EPJ, Abdurrahman2025, ScampsPRC2013, Regnier2019PRC, Simenel2010PRL, ScampsHashimoto2017PRC, Regnier2018PRC}. We have examined this limit as well and observe clear differences in the transfer probabilities between central and peripheral collisions, emphasizing that a meaningful comparison with experiment requires calculations at the appropriate impact parameter.

Our computational framework employs the PC-PK1 relativistic density functional \cite{ZhaoPRC2010}. Static calculations use a symmetric 3D lattice of $N_x\times N_y\times N_z = 24\times 24\times 24$ points with a 1 fm mesh spacing, while dynamical time evolution is computed on a larger, asymmetric lattice of $60\times 24\times 30$ points, with the same spacing. Pairing correlations are treated using a cutoff function from Ref.~\cite{RenPRC2022}, with pairing strengths ($G_n =-0.31$ MeV for neutrons, $G_p = -0.35$ MeV for protons in $^{96}{\rm Zr}$) determined from empirical gaps via the three-point odd-even mass formula. 

\begin{figure*}[!htbp]
	\centering
	\includegraphics[width=\textwidth]{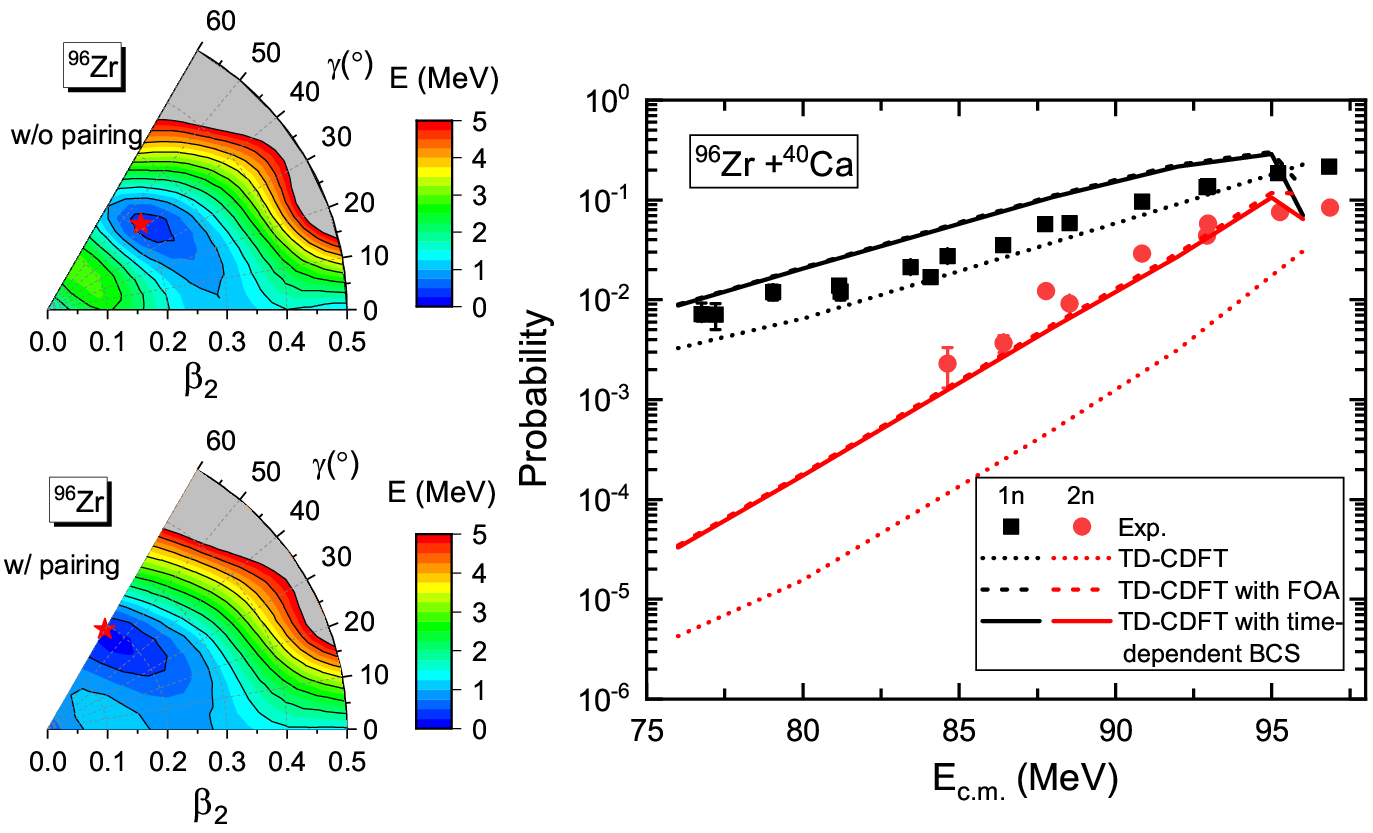}
	\caption{(Color online) One- and two-neutron transfer probabilities as a function of center-of-mass energy for the $^{96}{\rm Zr}+{}^{40}{\rm Ca}$ reaction at a fixed orientation (A1 configuration in ~Fig.~\ref{fig2}). The solid squares and circles are experimental values from Ref. ~\cite{Corradi2011PRC}. The dotted lines are calculated using TD-CDFT. The solid and dashed lines are obtained by TD-CDFT with time-dependent BCS approximation and with frozen occupation approximation (FOA), respectively. The potential energy surfaces of $^{96}{\rm Zr}$ calculated using CDFT without/with pairing correlations are shown in the left panel, with the equilibrium minimum denoted by the star.}
	\label{fig1}
\end{figure*}

The Dirac equation (Eq.\eqref{eq: Dirac equation}) is solved using a predictor-corrector method with a time step of 0.2 fm/c. In the initial configuration, the projectile and target nuclei are placed on the $x$-axis with an initial separation of 24 fm and are assumed to move along Rutherford trajectories. The time evolution is halted when two well-separated fragments are formed (center-of-mass distance $> 24$ fm) or when a composite system indicative of fusion is identified. Finally, particle number projection is performed by evaluating the integral in Eq.\eqref{eq: pnp_operator} using the trapezoidal method.

\section{Results and discussion}\label{sec4}

\begin{figure*}[!htbp]
	\centering
	\includegraphics[width=\textwidth]{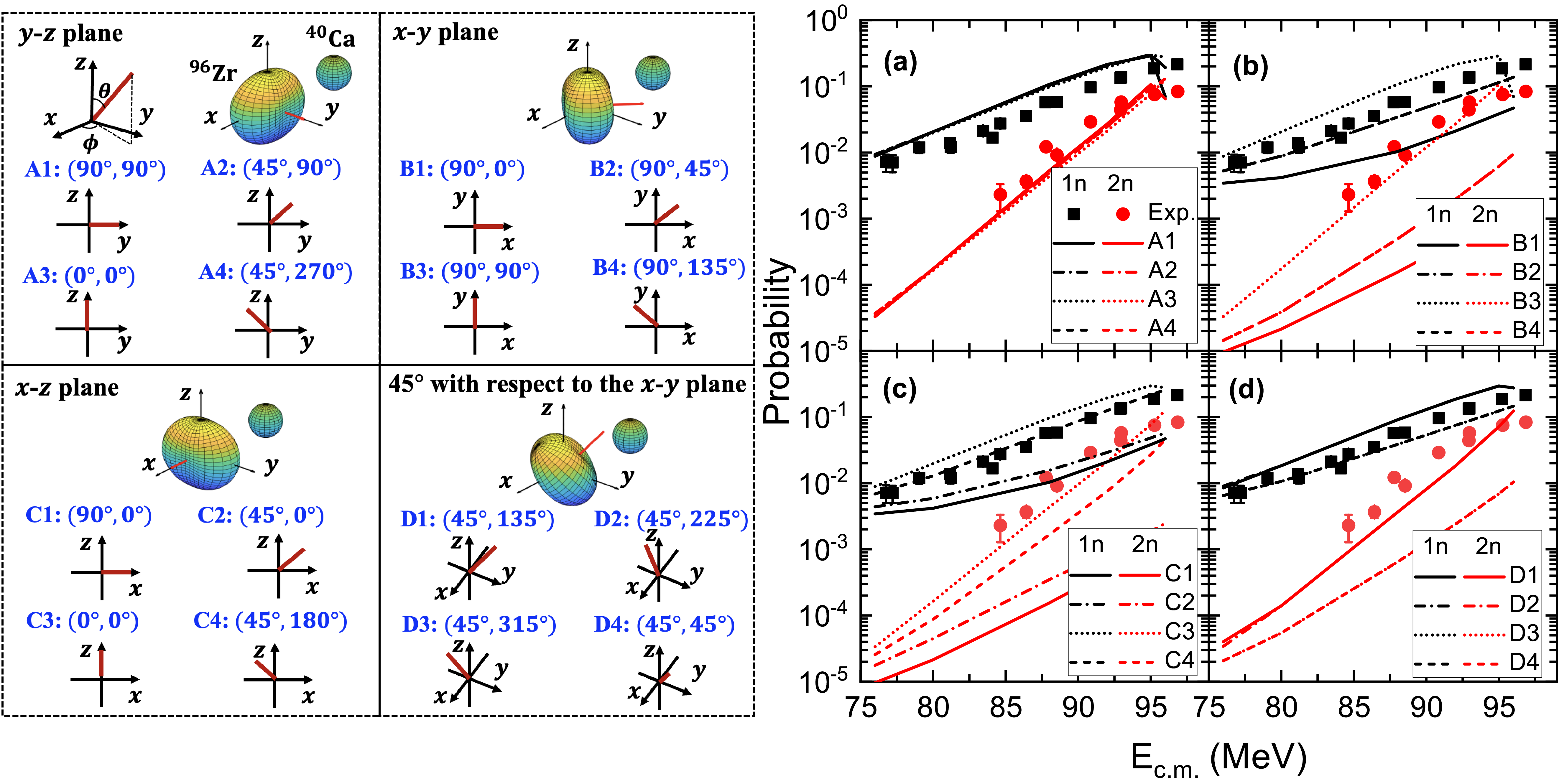}
	\caption{(Color online) One- and two-neutron transfer probabilities as a function of center-of-mass energy for the reaction $^{96}{\rm Zr}+{}^{40}{\rm Ca}$, calculated using TD-CDFT with dynamical pairing correlations. Results are shown for 16 orientations of the oblate $^{96}{\rm Zr}$ nucleus, comprising 13 distinct and 3 repeated configurations, defined schematically by the azimuthal angle $\phi$ and polar angle $\theta$ on the left: the symmetry axis lies in the (a) $y$-$z$, (b) $x$-$y$, and (c) $x$-$z$ planes, and (d) at a 45$^\circ$ angle relative to both the $z$ axis and the $x$-$y$ plane. Experimental data from Ref.~\cite{Corradi2011PRC} are shown as solid squares and circles. The different lines in each panel correspond to the specific orientations.}
	\label{fig2}
\end{figure*}

Sub-barrier one- and two-nucleon transfer has previously been studied within several variants of time-dependent density functional theory, including TDHF with particle-number projection~\cite{Simenel2010PRL}, TDHF+BCS~\cite{ScampsPRC2013, Scamps2015EPJ}, TDHFB for pair transfer between superfluid systems~\cite{ScampsHashimoto2017PRC, Regnier2018PRC, Regnier2019PRC}, and TDSLDA for the $^{40,48}{\rm Ca}+{}^{96}{\rm Zr}$ reaction~\cite{Abdurrahman2025}. These calculations did not include a scan over the orientation of a deformed reaction partner. The present calculations are performed at the experimentally relevant impact parameters $b\approx 2$-$3$ fm, retain the oblate deformation of $^{96}$Zr by sampling a systematic set of orientations, and include quantum fluctuations through the generalized TDGCM.

The time-dependent calculation is typically initialized with a Slater determinant composed of the ground states of $^{96}{\rm Zr}$ and ${}^{40}{\rm Ca}$, obtained from three-dimensional (3D) lattice CDFT \cite{Ren2017PRC, Zhang2022PRC, Zhang2023IJMPE, Ren2020PRC, Li2020PRC}. While the mean-field equilibrium state of ${}^{40}{\rm Ca}$ is spherical, $^{96}{\rm Zr}$ has a more complex structure. The left panel of Fig.~\ref{fig1} shows the potential energy surfaces (PES) for $^{96}{\rm Zr}$ with and without pairing correlations. The stars denote the mean-field equilibrium. Notably, when pairing correlations are included, the PES softens, and the quadrupole deformation of the ground state shifts from ($\beta=0.2, \gamma=47^{\circ}$) to ($\beta=0.2, \gamma=60^{\circ}$). 

In collisions involving deformed nuclei, the initial orientation is not unique. For the present analysis, a single orientation is selected, with a comprehensive discussion of orientation effects to follow later. In this configuration, the collision takes place in the $x$-$z$ plane, with the symmetry axis of the oblate $^{96}$Zr nucleus aligned along the $y$-axis.

The right panel of Fig.~\ref{fig1} shows the one- and two-neutron transfer probabilities as a function of center-of-mass energy for the $^{96}{\rm Zr}+{}^{40}{\rm Ca}$ reaction at a fixed orientation (A1 configuration in ~Fig.~\ref{fig2}). Solid squares and circles represent experimental values from Ref. \cite{Corradi2011PRC}. The dotted lines are the results from TD-CDFT, while the solid lines include the time-dependent BCS approximation to account for pairing correlations. The inclusion of pairing enhances both one- and two-neutron transfer probabilities. This enhancement can be partly attributed to the greater elongation of $^{96}$Zr along the $x$-axis in this orientation, which facilitates nucleon transfer (see further discussion below). 

For comparison, results calculated using the frozen occupation approximation (FOA) are shown as dashed lines. The transfer probabilities from TD-CDFT with FOA are generally consistent with those from the time-dependent BCS approach, except at the higher energy of $E_{\text{c.m.}}=96$ MeV. This consistency indicates that dynamical changes in occupation probabilities are strongly suppressed at energies well below the Coulomb barrier, whose height is $E_{\text{c.m.}} = 96.5$ MeV. At $E_{\text{c.m.}}=96$ MeV, reaction channels for multinucleon transfer begin to open, resulting in a decrease in the probabilities for one- and two-nucleon transfer.

Given the oblate deformation of $^{96}{\rm Zr}$ (with $\beta=0.2$), we selected a comprehensive set of orientations to assess their impact on one- and two-neutron transfer probabilities. As illustrated in the left panel of Fig.~\ref{fig2}, the symmetry axis of $^{96}{\rm Zr}$ lies in the $y$-$z$ plane (a), the $x$-$y$ plane (b), the $x$-$z$ plane (c), and at a 45$^\circ$ angle relative to both the $z$-axis and $x$-$y$ plane (d), resulting in 13 distinct orientations. The azimuthal angle $\phi$ is defined as the angle between the projection of the symmetry axis in the upper half-space ($z>0$) onto the $x$-$y$ plane and the positive $x$-axis (ranging from $0^\circ$ to $360^\circ$), while the polar angle $\theta$ is defined as the angle between this axis and the positive $z$-axis (ranging from $0^\circ$ to $90^\circ$). The corresponding energy-dependent transfer probabilities for each are displayed in the four right panels of Fig.\ref{fig2}. 

The initial collision geometry inherits specific symmetries from the spherical $^{40}$Ca and axially deformed $^{96}$Zr nuclei. Consequently, certain distinct orientations produce identical transfer probabilities. For example, in Fig.\ref{fig2} (a), the two orientations with the symmetry axis at 45$^\circ$ with respect to the $y$ and $z$ axes yield the same result, as indicated by the dash-dotted and dashed lines. Similar degeneracies are observed in Figs.~\ref{fig2} (b) and ~\ref{fig2} (d). Despite these symmetries, the results reveal that the transfer probabilities are highly sensitive to orientation, varying by several orders of magnitude across the different configurations. 

The transfer probabilities are maximized when the symmetry axis of $^{96}$Zr aligns with the $y$-axis (solid lines in Figs.~\ref{fig2} (a) and \ref{fig2} (b)), a configuration that presents the greatest elongation of the nucleus along the $x$-axis. Conversely, the probabilities are minimized when the symmetry axis aligns with the $x$-axis (dotted lines in Figs. \ref{fig2} (b) and \ref{fig2} (c)), resulting in minimal elongation along the $x$-axis. This correlation between elongation along the collision axis and transfer probability is consistent across all other orientations. 

\begin{figure}[!htbp]
	\centering
	\includegraphics[width=\columnwidth]{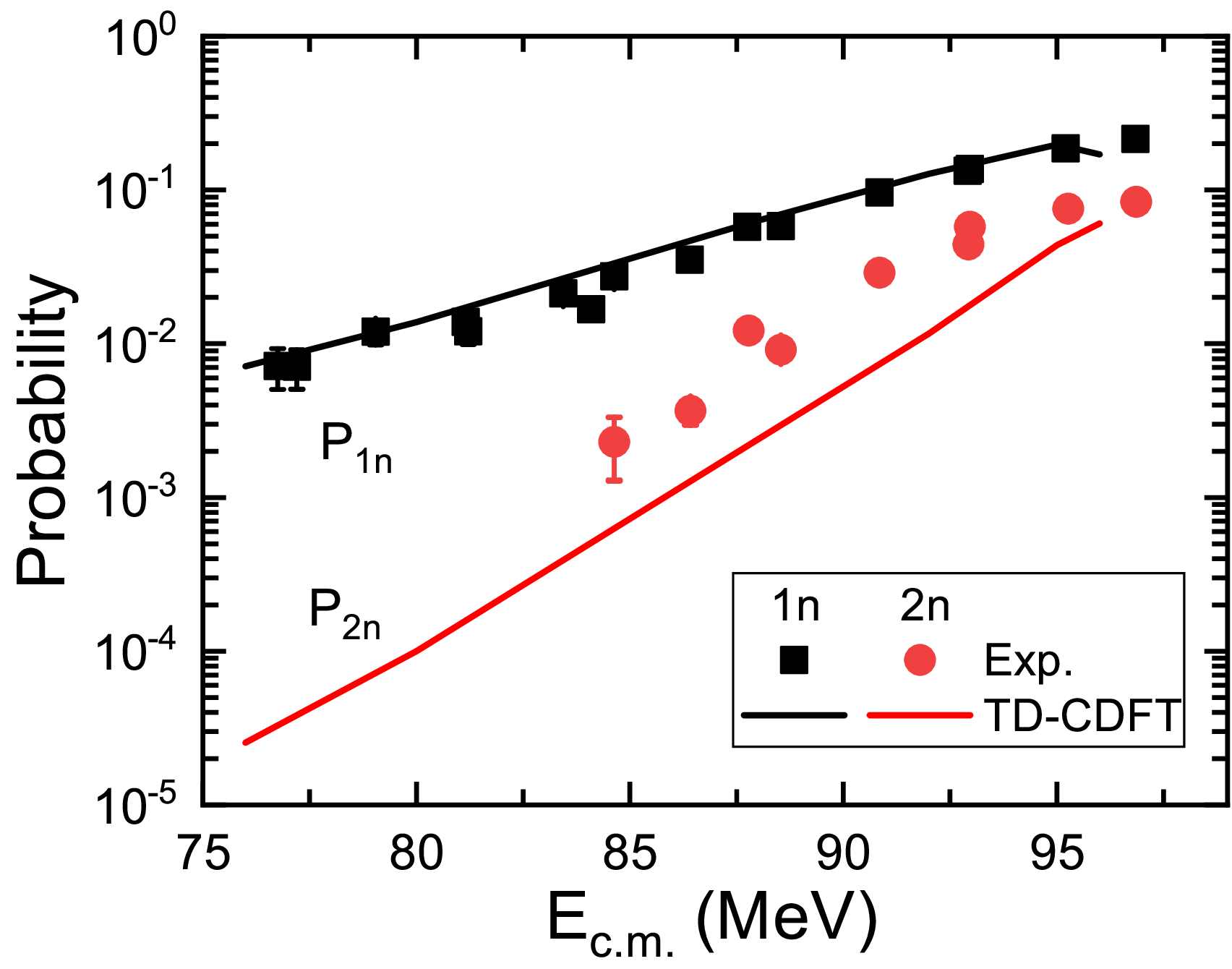}
	\caption{(Color online) One- and two-neutron transfer probabilities for the $^{96}{\rm Zr}+{}^{40}{\rm Ca}$ reaction, averaged over the 13 orientations calculated with TD-CDFT (including dynamical pairing correlations, see Fig.~\ref{fig2}), plotted as a function of center-of-mass energy.}
	\label{fig3}
\end{figure}

To facilitate comparison with experimental values, the calculated transfer probabilities from all 13 orientations were averaged with equal weight, as shown in Fig.~\ref{fig3}. The TD-CDFT approach with the time-dependent BCS approximation yields overall reasonable agreement with the data, particularly for one-neutron transfer, once this orientational averaging is performed. It is noteworthy, however, that the calculated two-neutron transfer probabilities remain lower than the experimental values at energies well below the barrier, a discrepancy where quantum fluctuations may play a significant role. To investigate this possibility, we have employed the generalized TDGCM framework.

\begin{figure}[!htbp]
	\centering
	\includegraphics[width=\columnwidth]{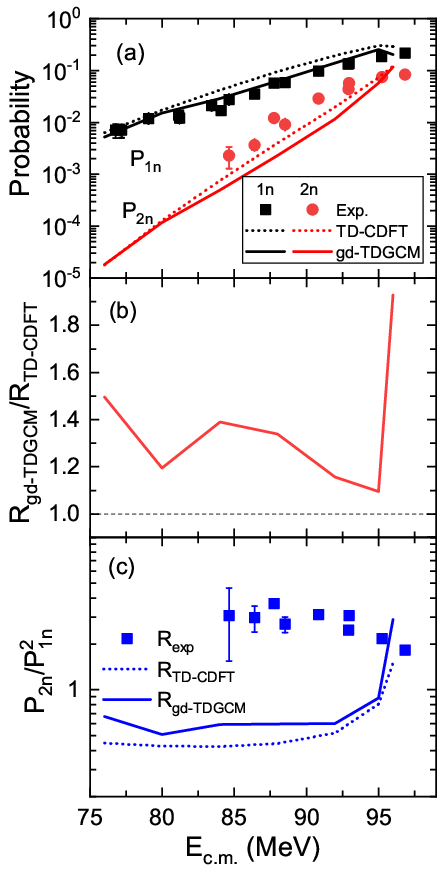}
	\caption{(Color online) (a) One- and two-neutron transfer probabilities for the $^{96}{\rm Zr}+{}^{40}{\rm Ca}$ reaction, calculated by TD-CDFT and generalized TDGCM (with dynamic pairing correlations). Both methods start from the initial state with the $^{96}$Zr symmetry axis aligned along the $z$-axis $(\theta, \phi)=(0^\circ, 0^\circ)$. The generalized TDGCM evolution employs five TD-CDFT generator states with initial orientations at  $(\theta, \phi)=(0^\circ, 0^\circ),(15^\circ,0^\circ),(30^\circ,0^\circ),(15^\circ,180^\circ),(30^\circ,180^\circ)$. (b) Enhancement factor $R_{\text{gd-TDGCM}}/R_{\text{TD-CDFT}}$, where $R=(P_{2n}/P_{1n}^2)$ is evaluated within the generalized TDGCM and TD-CDFT frameworks, respectively. (c) $R_{\text{TD-CDFT}}$ (dotted line), $R_{\text{gd-TDGCM}}$ (solid line), and $R_{\text{exp}}$ (squares with error bars where available) extracted from the data of Ref.~\cite{Corradi2011PRC}.}
	\label{fig4}
\end{figure}

Five TD-CDFT trajectories, starting from initial states with the symmetry axis of $^{96}$Zr at angles $(\theta, \phi)=(0^\circ, 0^\circ),(15^\circ,0^\circ),(30^\circ,0^\circ),(15^\circ,180^\circ),(30^\circ,180^\circ)$, were selected as generator states for the generalized TDGCM. The initial state for the generalized TDGCM calculation itself is identical to the TD-CDFT trajectory starting at $(0^{\rm{o}},0^{\rm{o}})$, where the symmetry axis is aligned with the $z$-axis. Within this framework, the $x$-neutron transfer probability for the generalized TDGCM is given by:
\begin{equation}
P^{\rm{TDGCM}}_{xn}= \sum_{\bm{q}} |g_{\bm{q}}|^2 P_{xn,\bm{q}}^{\text{TD-CDFT}},
\end{equation}
where $P_{xn,\bm{q}}^{\text{TD-CDFT}}$ is the $x$-neutron transfer probability of the corresponding TD-CDFT trajectory. Figure \ref{fig4} shows the one- and two-neutron transfer probabilities as a function of center-of-mass energy for the $^{96}{\rm Zr}+{}^{40}{\rm Ca}$ reaction, calculated using the generalized TDGCM with dynamic pairing correlations (solid line). For comparison, the results from a standard TD-CDFT calculation starting from the same $(0^{\rm{o}},0^{\rm{o}})$ initial state are shown as a dashed line. 

For one-neutron transfer, the generalized TDGCM yields smaller probabilities than TD-CDFT, resulting in closer agreement with the experimental data. For two-neutron transfer, however, the generalized TDGCM also produces smaller probabilities than TD-CDFT, but these are further from the experimental data in the incident energy range of 80 to 95 MeV.
Furthermore, the ratio $R=(P_{2n}/P_{1n}^2)$ is evaluated within both generalized TDGCM and TD-CDFT, which characterizes deviations from an uncorrelated sequential transfer mechanism. Figure 4(b) displays the enhancement factor $R_{\text{gd-TDGCM}}/R_{\text{TD-CDFT}}$, where $R=(P_{2n}/P_{1n}^2)$ is evaluated within the generalized TDGCM and TD-CDFT frameworks, respectively. The results show that the inclusion of quantum fluctuations systematically enhances this ratio, indicating increased correlations between the transferred neutrons. This suggests that quantum fluctuations beyond the mean-field approximation also make an essential contribution to the correlated neutron transfer. However, since we are considering sub-barrier energies, both methods based on TD-DFT predict values for the ratio $R$ that underestimate the data, as shown in Fig.~\ref{fig4}(c). It is only when $E_{\text{c.m.}}$ approaches the Coulomb barrier that the calculated values for $R$ reproduce the data.

\section{Summary}\label{sec5}

The sub-barrier transfer dynamics of the $^{96}{\rm Zr}+{}^{40}{\rm Ca}$ reaction were investigated using a framework based on time-dependent covariant density functional theory, with pairing correlations incorporated via the time-dependent BCS approximation. Furthermore, the neutron transfer process exhibits extreme sensitivity to the orientation of the deformed $^{96}$Zr nucleus, with probabilities varying by several orders of magnitude.

To compare with experimental data, orientation-averaged probabilities were computed. While the averaged one-neutron transfer results agree well with data, the two-neutron transfer probabilities are suppressed below the barrier. We attribute this suppression to the semiclassical nature of TD-CDFT, which lacks quantum fluctuations. To quantify this effect, we employed the generalized time-dependent generator coordinate method. The results indicate that quantum fluctuations enhance the correlations between transferred neutrons, as reflected in the ratio $R=(P_{2n}/P_{1n}^2)$.

However, our current TDGCM implementation does not include all relevant quantum fluctuations, particularly those associated with the time evolution of the generator states, due to the absence of corresponding equations of motion from a time-dependent variational principle. A comprehensive treatment of these additional fluctuations is a crucial objective for future work to achieve a fully quantitative understanding of sub-barrier transfer dynamics.

\section*{Acknowledgments}
The authors are grateful to S. G. Zhou for useful discussions. This work is supported by the National Key Research and Development Program of China (Contract No. 2024YFE0109800, No. 2023YFA1606500, No. 2024YFE0109803 and No. 2024YFA1612600),  and the CAS Strategic Priority Research Program (Grant No. XDB1550102, XDB34010100, and No. XDB0920000), the National Natural Science Foundation of China (Grants No. 12447101, No. 12375118, No. 12435008, No. W2412043, No. 12435006, No. 12475117, No. 12141501, and No. 11935003). 
This work has been supported in part by the High-End Foreign Experts Plan of China, the National Key Laboratory of Neutron Science and Technology (Grant No. NST202401016). 
This work has been supported in part by the project “Implementation of cutting-edge research and its application as part of the Scientific Center of Excellence for Quantum and Complex Systems, and Representations of Lie Algebras”, Grant No. PK.1.1.10.0004, co-financed by the European Union through the European Regional Development Fund - Competitiveness and Cohesion Programme 2021-2027 and by the Croatian Science Foundation under the project Dynamics of complex femtosystems (IP-2025-02-4214).
This work is supported by the Postdoctoral Fellowship Program and China Postdoctoral Science Foundation under Grant Number BX20250170.
The results described in this paper are obtained on the High-performance Computing Cluster of ITP-CAS and the Sc-Grid of the Supercomputing Center, Computer Network Information Center of the Chinese Academy of Sciences.

\section*{Data Availability}
The data that support the findings of this article are not publicly available. The data are available from the authors upon reasonable request.

%

\end{document}